\documentstyle[twoside,fleqn,espcrc2]{article}

\newcommand{\AmS}{{\protect\the\textfont2
  A\kern-.1667em\lower.5ex\hbox{M}\kern-.125emS}}

\title{High performance Beowulf computer for lattice QCD}

\author{Xiang-Qian Luo$^a$, E.B. Gregory$^a$, H.J. Xi\address{ 
Department of Physics, Zhongshan University, 
Guangzhou 510275, China}
J.C. Yang$^b$, Y.L. Wang$^b$, Y. Lin\address{Gosun Internet Information 
and Technology Development Co., Ltd, Guangzhou 510080, China}, 
and 
H.P. Ying\address{Zhejiang Institute of Modern Physics, Zhejiang University, Hangzhou 310027, China}
}

\begin{document}

\begin{abstract}
We describe the construction of a high performance parallel computer 
composed of PC components, 
as well as the performance test in lattice QCD.
\end{abstract}

\maketitle

\section{INTRODUCTION}

The ZhongShan University Computational Physics group's 
interests \cite{zsusite} cover such topics
as lattice QCD\cite{luo,luo2,gregory,Luo:2001id}
quantum instanton\cite{PLA} and quantum chaos\cite{PRL}. 
Most of these topics can be 
studied through Monte Carlo simulation, but can be quite costly in 
terms of computing power.  In order to do large scale 
numerical investigations of these topics, we required a corresponding 
development of our local computing resources.  

The demarcation 
between super computers and personal computers has been 
further blurred in recent years by the high speed and low price of modern 
CPUs and networking technology and the availability of low cost or free
software.  By combining these three elements - all readily 
available to the consumer - one can assemble a true super computer that is
within the budget of small research labs and businesses. 

  We document the construction and performance of a Beowulf cluster of PCs, configured 
to be capable of parallel processing.

\section{SYSTEM}

Our cluster consists of ten PC type computers\cite{Luo:2001mp}, each with two
Pentium III-500 CPUs inside.  The logic behind dual CPU machines is that one
can double the number of processors without the expense of additional, cases,
power supplies, motherboards, network cards, et cetera. Also, the inter-node 
communication speed is faster for each pair of processors in the same box
as compared to communication between separate computers. Each computer has 
an 8GB EIDE hard drive, 128 MB of memory, a 100Mbit/s Ethernet card, a simple
graphics card a floppy drive and a CDROM.  In practice the CDROM, the floppy 
drive, and even the graphics card could be 
considered extraneous,  as all interactions with the nodes could be done 
through the network.
One computer has a larger hard disk (20 GB), and a SCSI card for 
interaction with a tape drive.  For the entire cluster we have only one console
consisting of a keyboard, mouse and monitor.

A fast Ethernet switch handles the inter-node communication. The switch has 
24 ports so there is ample room for future expansion of the cluster to up to 
a total of 48 processors. Of course it is possible to link multiple switches
or use nodes with more that two processors, so the possibilities for a larger 
cluster are nearly limitless.

We have installed a Red Hat Linux 6.1 distribution.  It automatically 
supports dual CPU computers.  It is also able to support a 
Network File System (NFS), allowing all of the nodes in the cluster to share
hard disks, and a Network Information System (NIS), which standardizes the
usernames and passwords across the cluster.  

We can use the the cluster for parallel processing by using the message 
passing interface (MPI), 
a library of communication functions and programs 
that allow for communication between processes on different CPUs.
The programmer must design the parallel algorithm so that it appropriately
divides the task among the individual 
processors. He or she must then include message passing 
functions in the code which allow information to be sent and received by the 
various processors.

\section{PERFORMANCE}

\begin{table}
\begin{tabular}{|c|c|c|} 
\hline 
\raisebox{0pt}[12pt][6pt]{Machine} & 
\raisebox{0pt}[12pt][6pt]{$\mu$-sec/link} & 
\raisebox{0pt}[12pt][6pt]{MB/sec}\\
\hline
\raisebox{0pt}[12pt][6pt]{SX-4} & 
\raisebox{0pt}[12pt][6pt]{4.50} & 
\raisebox{0pt}[12pt][6pt]{45}\\

\raisebox{0pt}[12pt][6pt]{SR2201} & 
\raisebox{0pt}[12pt][6pt]{31.4} & 
\raisebox{0pt}[12pt][6pt]{28}\\
\raisebox{0pt}[12pt][6pt]{Cenju-3} & 
\raisebox{0pt}[12pt][6pt]{57.42} & 
\raisebox{0pt}[12pt][6pt]{8.1}\\
\raisebox{0pt}[12pt][6pt]{Paragon} & 
\raisebox{0pt}[12pt][6pt]{149} & 
\raisebox{0pt}[12pt][6pt]{9.0}\\
\raisebox{0pt}[12pt][6pt]{\bf ZSU Cluster} & 
\raisebox{0pt}[12pt][6pt]{\bf 3.98} & 
\raisebox{0pt}[12pt][6pt]{\bf 11.5}\\
\hline
\end{tabular}
\caption{Performance of MPI QCD benchmark \cite{hioki}.}\label{table1}
\label{tab:qcdim}
\end{table}

As we primarily developed the cluster for numerical simulations of 
lattice QCD, we have also performed a benchmark which specifically tests the 
performance in a parallel lattice QCD code.  
The algorithm can conveniently divide the lattice and assign the sections
to different processors. 

Hioki and Nakamura \cite{hioki}
provide comparison performance data on SX-4 (NEC), SR2201 (Hitachi),
 Cenju-3 (NEC) and  Paragon (Intel) machines.
Specifically, we compare the computing time per link update in microseconds 
per link and the inter-node communication speed in MB/sec. The link update is a
fundamental computational task within the QCD simulation and is therefore a 
useful standard.  The test was a simulation of improved pure gauge
action ($1\times 1$ plaquette and $1\times 2$ rectangle terms) on a 
$16^4$ lattice. In each case the simulation was run on 16 processors.  The 
results are summarized in Table \ref{tab:qcdim}.

A widely used QCD program is the MILC code \cite{milc}. 
It has timing routines provided so that 
one can use the parallelized conjugate gradient routine in the simulation 
as a benchmark.
Furthermore, as this code is very versatile and is designed to be run on a 
wide variety of computers and architectures.  This enables quantitative 
comparison of our cluster to both other clusters and commercial supercomputers.
In the MILC benchmark test we ran to a convergence tolerance of $10^{-5}$ per
site.  For consistency with benchmarks performed by others, we simulated 
Kogut-Susskind fermions.

We have run the 
benchmark test for different size lattices and different numbers of processors.
It is useful to look at how performance is affected by the number of CPUs, 
when the amount of data per CPU is held fixed, that is each CPU is 
responsible for a section of the lattice that has $L^4$ sites. For one CPU,
the size of the total lattice is $L^4$. For two CPUs it is $L^3\times 2L$.
For four CPUs the total lattice is $L^2 \times (2L)^2$; for eight CPUs, 
$L \times (2L)^3$, and for 16 CPUs the total size of the lattice is $(2L)^4$.

Note that the falloff in performance with increased number of CPUs is dramatic.
This is 
because inter-processor message passing is the slowest portion of this or any
MPI program and from two to sixteen CPUs, the amount of communication per
processor increases by a factor of four. 
 Table \ref{boundary} shows that for a lattice divided into $2^j$
hypercubes, each of size $L^4$, there will be $j$ directions in which the 
CPUs must pass data to their neighbors. The amount of communication each 
processor must perform is proportional to the amount of interface per 
processor. As this increases, per node performance decreases until $j=4$ and 
every lattice dimension has been divided (for a $d=4$ simulation). and the
per-processor performance should remain constant as more processors are added.
The shape of this decay is qualitatively consistent with $1/j$ falloff.

\begin{table}
\begin{center}
\begin{tabular}{|c|c|c|c|c|}
\hline
I.D.
&
H. 
& 
L.V.  
&
T.I. 
&
I./CPU
\\
\hline
$j$  &  $2^j$  &  $L^{4-j} \times (2L)^j$  & $2^jjL^{3}$ & $jL^{3}$\\
\hline
0 & 1 & $L^4$ & 0 & 0 \\
1 & 2 & $L^3 \times 2L $ & $2L^3$ &  $L^3$\\
2 & 4 & $L^2 \times (2L)^2 $ & $8L^3$ & $2L^3$\\
3 & 8 & $L \times (2L)^3 $ & $24L^3$  & $3L^3$\\
4 & 16 & $(2L)^4 $ & $64L^3$  & $4L^3$\\
\hline
\end{tabular}
\end{center}
\caption{ \label{boundary} Boundary sizes for division of a lattice
into 1,2,4,8 and 16 hypercubes of size $L^4$. Here 
I.D. stands for interface directions, 
H. for hypercubes (CPUs) 
L.V. for
lattice volume,
T.I. for total interface,  
and I./CPU for
interface/CPU respectively.
}
\end{table}

Of course there are other ways to divide a four-dimensional lattice. 
The goal of a particular simulation will dictate the geometry of the lattice 
and the therefore the most efficient way to divide it up (generally minimizing 
communication).  A four-CPU simulation using a $4L\times L^3$ lattice has 
the four hypercubic lattice sections lined up in a row (as opposed to in a 
$2\times 2$ square for a $L^2 \times (2L)^2$ lattice) and has the same amount 
of communication per CPU as does the $L^3\times 2L$
two-CPU simulation.  In a benchmark test the per-CPU performance was
comparable to the performance in the two-CPU test.

For a single processor, there is a general decrease in performance as $L$ 
increases, as shown in Table \ref{one_cpu}.
This is well explained in \cite{gottlieb} as due to the larger matrix size
using more space outside of the cache memory, causing slower access time to 
the data. 

\begin{table}
\begin{center}
\begin{tabular}{|c|c|}
\hline
$L$ & single processor speed (Mflops)\\
\hline
4 & 161.5\\
6 & 103.2\\
8 & 78.6\\
10 & 76.4\\
12 & 73.9\\
14 & 75.9\\
\hline
\end{tabular}
\end{center}
\caption{ \label{one_cpu} Single CPU performance of MILC code.}
\end{table}

For multiple CPUs there is in performance improvement as $L$ 
is increased\cite{Luo:2001mp}.
The explanation for this is that the communication bandwidth is not constant 
with respect to message size. For very small 
message sizes, the bandwidth is very poor. It is only with messages of around 
10kB or greater that the bandwidth reaches the full potential of the fast 
Ethernet hardware, nearly 100Mbit/sec. With a larger $L$, the size of the
messages is also, improving the communication efficiency. The inter-node
communication latency for our system is 102$\mu s$. As inter-node 
communication is the slowest part a parallel program this far outways the 
effect of cache misses.

To summarize,
a parallel cluster of PC type computers is an 
economical way to build a powerful computing resource for academic purposes.
On an MPI QCD benchmark simulation it compares favorably with other MPI 
platforms. The price/performance ratio\cite{Luo:2001mp} is $\$7/Mflop$.
It is drastically cheaper than commercial supercomputers 
for the same amount of processing speed. 
It is particularly suitable for 
developing research groups in countries where funding for pure research is 
more scarce. 
We have been doing large scale calculations of hadron and glueball spectrum\cite{Spectrum}.

X.Q.L. is supported by 
National Science Fund for Distinguished Young Scholars (19825117),
Guangdong Provincial Natural Science Foundation (990212), 
Ministry of Education, Gosun 
Ltd., 
and the Foundation of Zhongshan Univ. Advanced Research Center. 
We also thank S. Hioki and S. Gottlieb for their benchmark codes.

\end{document}